# Prediction for structure stability and ultrahigh hydrogen evolution performance of monolayer 2H-CrS$_2$

Feng Sun[1], Aijun Hong[1*], Wenda Zhou[1], Cailei Yuan[1*], Wei Zhang[2]

*Correspondence and requests for materials should be addressed to A.J.H. (email: haj@jxnu.edu.cn) and C.L.Y. (email: clyuan@jxnu.edu.cn).

1 Jiangxi Key Laboratory of Nanomaterials and Sensors, School of Physics, Communication and Electronics, Jiangxi Normal University, Nanchang 330022, China

2 State Key Laboratory of Hydroscience and Engineering, Department of Energy and Power Engineering, Tsinghua University, Beijing 100084, China

**ABSTRACT:** By a combination of the first-principles calculations and climbing image nudged elastic band method (ciNEB) we investigate structure stabilities and hydrogen evolution reaction (HER) performance of monolayer 2H-CrS$_2$. The results suggest the free energy for the Volmer reaction in the monolayer 2H-CrS$_2$ with S vacancy is 0.07 eV, comparable with Pt-based catalyst, and HER on the surface of the monolayer is prone to the Volmer-Heyrovsky mechanism with no energy barrier. We propose that high HER performance stems from the reduction of the energy level of d-band center. Additionally, the S vacancy leads to defect states in the middle of electronic bandgap and the reduction of potential barrier between the S atom layer and the vacuum, which is conducive to improve HER performance.

## 1. Introduction

In the past decades, two-dimensional (2D) layered materials, due to special physical and chemical properties, have received unprecedented attention [1-4]. Especially, transition-metal dichalcogenides (TMDs) [5-7], as an important member of the 2D material family, present promising applications in energy field such as battery, solar cell and especially hydrogen evolution reaction (HER) [7-10]. It is worth mentioning that $MoS_2$ has been regarded as a likely substitute of costly Pt-based catalyst for HER because of high elemental abundance and high catalytic performance. For instance, the doped $MoS_2$ in the 2H structure has low onset potential of ~0.13 eV and slight Tafel slope of 49 mV dec$^{-1}$ [8].

However, there is little attention on 2D chromium disulfide ($CrS_2$) because it and its bulk counterpart have not been synthesized experimentally. Thankfully, recent work [9] claimed that monolayer $CrS_2$ with multiphase coexisting could be successfully prepared via the chemical vapor deposition (CVD). On the other band, it has been predicted theoretically that the ground state of $CrS_2$ monolayer possesses the 2H phase. These make us believe that 2H-$CrS_2$ monolayer can be successfully prepared through the special experimental method. Thus, it is of prospective signification to explore HER performance and mechanism of 2H-$CrS_2$ monolayer theoretically for improved HER catalytic effects. At present, there are two common and effective strategies for the improvement of HER performance. One is to increase density of catalytic active sites and the other is to enhance electrical conductivity. Defect engineering can realize killing two birds with one stone because effective control of exterior or intrinsic defect could provide a large number of active sites and high carrier density thus leading to high electrical conductivity. Accordingly, it is meaningful and necessary to study the HER catalytic effects in the 2H-$CrS_2$ monolayer by defect engineering.

In this work, we employ density functional theory (DFT) combined with ciNEB to explore mechanical, dynamical, and thermal stabilities of 2H-$CrS_2$ monolayer with/without intrinsic defect. The results show both Cr vacancy ($V_{Cr}$) and S vacancy ($V_S$) in the imperfect monolayer have low Volmer reaction free energies of 0.11 eV and 0.07 eV, comparable with Pt-based catalyst. The calculation results for the formation energy present $V_S$ he 2H-CrS$_2$ monolayer is easier to form. Then, we explore reaction pathways for Heyrovsky and Tafel reactions of the 2H-CrS$_2$ monolayer with $V_S$ using the ciNEB method, and conclude its HER belongs to the Volmer−Heyrovsky mechanism.

## 2. Calculation details and models

We add vacuum of 15 Å thick to all computational structures (4×4×1 supercells) for shielding the interaction between the periodic images and then geometrically optimize the structures by using VASP code [10-12]. In all DFT calculations, we adopt the generalized gradient approximation (GGA) and Perdew−Burke−Ernzerhof functional (PBE) [10, 12] as the exchange correlation potential with a cutoff energy of 500 eV. For the geometrical optimization calculation, the total energy and force criteria are set to less than $1×10^{-4}$ eV and 0.03eV/Å, and a $k$-mesh of 5×5×1 is used. However, we increase $k$-mesh density to 11×11×1 for the calculation of electronic densities of states (DOS). The ciNEB method [13] is employed to investigate HER mechanism. In this process, the force criteria are set to 0.04 eV/Å, and nine images are inserted between the initial and final states (IS and FS) to locate the reaction pathways and the transition states. Phonon structures are obtained by using Phonopy code [14], where the density functional perturbation method (DFPT) [15] and a 5×5×1 supercell are employed. Ab initio molecular dynamics (MD) simulation based on the canonical ensemble (NVT) [16, 17] is employ to verify structural thermal stability.

## 3. Results and discussion

**A Stabilities of 2H-CrS$_2$ monolayer**

In this section, we use elastic constants, phonon spectrum, and ab initio molecular dynamics (MD) simulation to confirm mechanical, dynamical, and thermal stabilities of 2H-CrS$_2$ monolayer. The monolayer with the vacuum can be seen as three dimensional (3D) periodic image, belonging to the hexagonal crystal. Therefore, there are six independent elastic constants $C_{11}$, $C_{12}$, $C_{13}$, $C_{14}$, $C_{33}$ and $C_{44}$, as summarized in Table SI. The elastic constants fulfill mechanical stability criteria [18], i.e., $C_{44} > 0$, $C_{66} >$

0, $C_{11} > |C_{12}|$, and $(C_{11} + 2C_{12})C_{33} > C_{13}^2$.

Admittedly, the elastic property of 2H-CrS$_2$ monolayer can also be described by the 3×3 elastic stiffness matrix [19]:

$$\begin{pmatrix} C_{11} & C_{12} & 0 \\ C_{12} & C_{22} & 0 \\ 0 & 0 & C_{66} \end{pmatrix}, \qquad (1)$$

herein, $C_{66}$ is equal to $(C_{11}-C_{12})/2$. According to the Born criteria [20]: $C_{11} > 0$ and $C_{11}-C_{12} > 0$, which also ensures the 2H-CrS$_2$ monolayer is mechanically stable.

In Fig. 1 (a) and (b), we present the phonon dispersion and phonon densities of states (PDOS) calculated within DFPT. The primitive cell of the monolayer 2H-CrS$_2$ has three atoms and thus possesses nine phonon modes. Clearly, there are no imaginary frequencies, further verifying its dynamical stability. PDOS in low frequency and high frequency ranges come from the joint contribution of Cr and S elements, although S element has a main contribution to the middle frequency range.

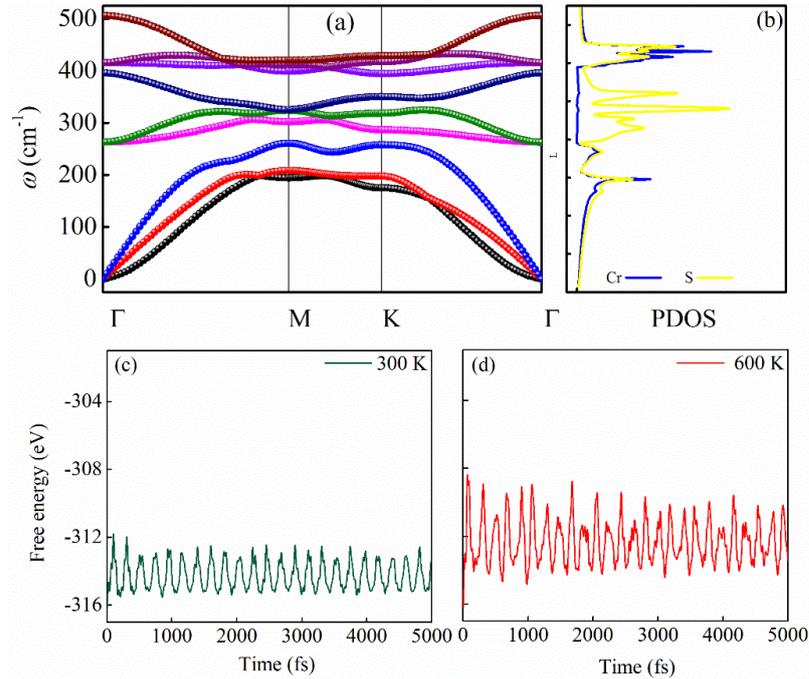

FIG. 1. (a) Calculated phonon dispersion and (b) phonon densities of states (PDOS) for the 2H-CrS$_2$ monolayer. Free energy as a function of time for the 2H-CrS$_2$ monolayer (c) at 300 K and (d) at 600 K.

Subsequently, to determine thermal stability, we carry out ab-initio molecular dynamics (MD) simulations of 2H-CrS$_2$ monolayer at 300 and 600 K, as shown in Fig. 1 (c) and (d). We find the total energy remains small fluctuation and the structure has no obvious change, although the positions of atoms have slight movements. Therefore, monolayer 2H-CrS$_2$ is thermally stable in a wide temperature range from room temperature to medium temperature [21].

**B HER performance and mechanism**

Before this part of the discussion, it is necessary to deliberate about the electronic structure information of the monolayer 2H-CrS$_2$ because good catalysts usually share unique electronic structure characteristics. For example, the famous d-band center theory [22] proposed that the energy level of d-band center ($\varepsilon_d$) is critical to catalytic activity. It is with respect to the density of states (DOS) of d electron $D_d$ (E) and energy $E$,

$$\varepsilon_d = \frac{\int_{-\infty}^{+\infty} ED(E)dE}{\int_{-\infty}^{+\infty} D(E)dE}. \tag{2}$$

Fig. 2 shows calculated electronic band structures and densities of states (DOS) of the perfect monolayer 2H-CrS$_2$ (P) and imperfect monolayers (V$_{Cr}$ and V$_S$) systems. Obviously, the introduction of Cr or S vacancy induces defect states in the middle of bandgap consisting of strong hybridization of Cr-d and S-s electrons, indicating the vacancy can strength the coupling of atoms around the vacancy. Moreover, the hybridization affects the DOS of Cr-d states and thus leads to the change of the energy level of d-band center $\varepsilon_d$. The $\varepsilon_d$ values for the V$_{Cr}$ and V$_S$ systems relative to the Fermi level are -1.08 eV and -1.02 eV that are lower than -0.87 eV of the P system. Previous work [23] clearly demonstrates the downshift of $\varepsilon_d$ makes the adsorption energy of H down and meanwhile is good for desorption of H from the catalyst surface for HER effects.

On the other hand, the vacancy changes the total potential distributions in Fig. 2(g), providing a possibly opportunity to improve HER performance. All the total potential curves like double-hump shape, corresponding to the sandwich structure of 2H-$CrS_2$ monolayer. Two S atomic layers have relatively low total potentials and Cr atomic layer's is high. Thus, to participate in the HER, an electron in the bottom S atomic layer needs to overcome the potential wells to the other sulfur layer, and then crosses the last potential well to reach the surface. The depth of the last potential well in $V_S$ system is 2 eV smaller than that for P and $V_{Cr}$ systems, which supports high electron tunneling probability and thus can improve HER performance.

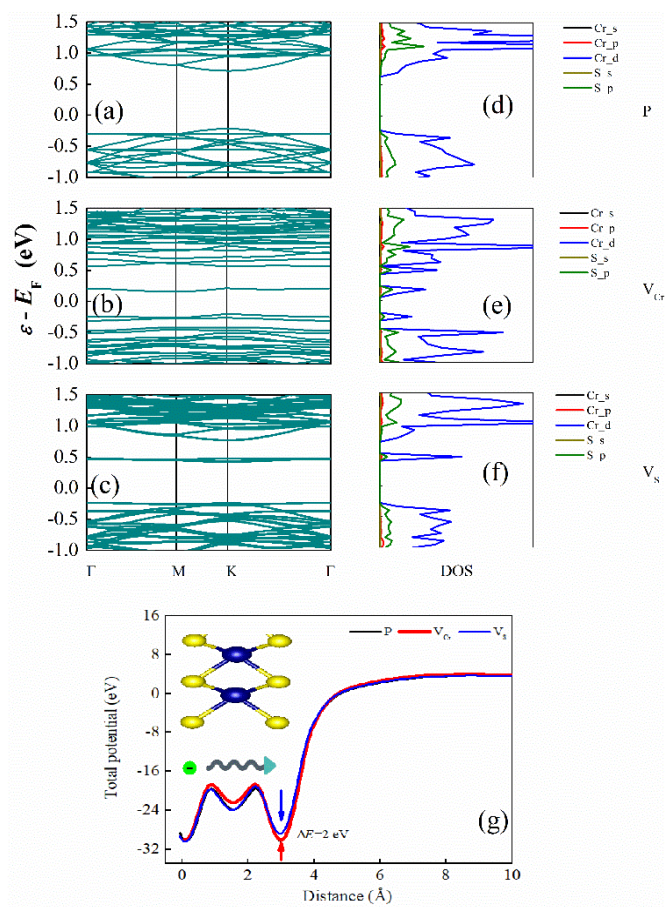

FIG. 2. Calculated electronic band structures and DOS for P [(a) and (d)], $V_{Cr}$ [(b) and (e)] and $V_S$ [(c) and (f)]. Total potential distributions of the P, $V_{Cr}$ and $V_S$ systems along the c axis.

The HER, as a tanglesome electrochemical process, usually happens on the surface of catalyst. At present, there are two recognized mechanisms for HER in acid solution [24]: one is Volmer-Heyrovsky reaction and the other is Volmer-Tafel reaction. The former reaction process includes two steps. First electrochemical hydrogen adsorption, namely, Volmer reaction takes place,

$$H^+ + e^- + X = H^*\text{-}X, \tag{3}$$

and then electrochemical desorption (Heyrovsky reaction) occurs, as follows:

$$H^*\text{-}X + e^- + H^+ = X + H_2. \tag{4}$$

However, the Volmer-Tafel reaction possesses the different way of H desorption (Tafel reaction),

$$H^*\text{-}X\text{-}H^* = X + H_2, \tag{5}$$

herein, $H^*$ is a hydrogen atom adsorbed on the active site of catalyst surface. In a word, the difference between the two mechanisms lies in the different ways of H desorption. Adsorption and desorption of H atom on the catalyst surface have competitive relation. Thus, the balance between adsorption and desorption is important for a high-efficiency catalyst. According to the Sabatier's principle [4], to achieve the balance, the free energy differences ($\Delta G = G_P - G_R$) between the free energies of product and reactant ($G_P$ and $G_R$) in each reaction, namely, the reaction free energies should be near zero. One can define the free energy of product and reactant ($G_P$ and $G_R$) as the following form

$$G = E_{ele} + E_{vib} - TS_{vib} + pV \tag{6}$$

where $E_{ele}$ is the total energy associated with electrons, $E_{vib}$ the vibrational energy, $T$ absolute temperature, $S$ the vibrational entropy, $p$ the pressure, $V$ the volume. Both the vibrational energy and entropy of an $N$-atomic system can be attained by the partition function:

$$Z = \sum_{l=1}^{3N} \sum_{n=0}^{\infty} e^{(n+\frac{1}{2})\beta\hbar\omega_l} \tag{7}$$

where $\beta = 1/kT$ ($k$: Boltzmann constant) and the $\omega_l$ has $3N$ vibrational modes. Then, the vibrational energy and entropy are given by

$$E_{vib} = -\frac{\partial}{\partial \beta} \ln Z \tag{8}$$

and

$$S_{vib} = k(\ln Z + \beta E_{vib}) \tag{9}$$

Using Eq. (7) in Eqs. (8) and (9), one can finally write

$$E_{vib} = \sum_{l=1}^{3N} \hbar\omega_l \left(\frac{1}{2} + \frac{1}{e^{\beta\hbar\omega_l}-1}\right) \tag{10}$$

and

$$S_{vib} = \sum_{l=1}^{3N} k \left[\frac{\beta\hbar\omega_l}{e^{\beta\hbar\omega_l}-1} - \ln(1-e^{-\beta\hbar\omega_l})\right]. \tag{11}$$

When the temperature approaches to 0 K, the vibrational energy is written as

$$E_{vib} = \sum_{l=1}^{3N} \frac{1}{2}\hbar\omega_l, \tag{12}$$

which is also called zero point energy ($E_{ZPE}$). Table SII illustrates $E_{vib}$ and $TS_{vib}$ of one adsorbed H on the surfaces of all the systems. These values for all the systems except the perfect system with H-adsorption on the top of Cr ($P_{Cr}$), are comparable with $E_{vib}$ (0.14 eV) and $TS_{vib}$ (0.2 eV) for one H atom of the $H_2$ gas molecule under standard conditions.

Finally, the reaction free energies of Volmer ($\Delta G_{nH}^V$), Heyrovsky ($\Delta G^H$) and Tafel ($\Delta G^T$) reactions can be computed using the following formulas

$$\Delta G_{nH}^V = \Delta E_{nH} + \Delta E_{vib} - T\Delta S_{vib}, \tag{13}$$

$$\Delta G^H = -\Delta G_{nH}^V, \tag{14}$$

$$\Delta G^T = -\Delta G_{nH}^V - \Delta G_{(n+1)H}^V, \tag{15}$$

where $\Delta E_{nH}$, $\Delta E_{vib}$ and $\Delta S_{vib}$ are the adsorption energy, the vibration energy and entropy differences in the adsorption reaction. As known, the Volmer reaction is the foremost step in the HER. Therefore, we first put focus on the Volmer reaction. All the possible H-adsorption sites in each surface have been taken into account, and the typical adsorption configurations with the lowest total energy are displayed in Fig. 3, where $P_{Cr}$ and $P_S$ stand for the P system with H-adsorption on the top of Cr and with H-adsorption on the top of S, respectively. The length of S-H in $P_S$ is 1.36 Å that is shorter than the Cr-H length of 1.64 Å in $P_{Cr}$. The $\Delta G_H^V$ for $P_S$ is 1.35 eV that is smaller than 1.76 eV for $P_{Cr}$, suggesting they are suitable for hydrogen adsorption but desorption. For the $V_{cr}$ system, H atom chemically binds to a prominent S atom with the S-H length of 1.37 Å. However, the Cr-H length reaches up to 3 Å, implying weak interaction between Cr and H atoms. For the $V_S$ system, the distance between H atom and the three nearest neighbor Cr atoms is 1.92 Å, and that between H atom and the nearest neighbor S atom is 2.4 Å, implying relatively weak H-S bond. The $\Delta G_H^V$ for both $V_S$ and $V_{Cr}$ are very close to 0 eV, comparable with Pt-based catalyst. This fully illustrates that the Cr and S vacancies as two types of intrinsic defects are efficient active sites in the imperfect 2H-$CrS_2$ monolayer.

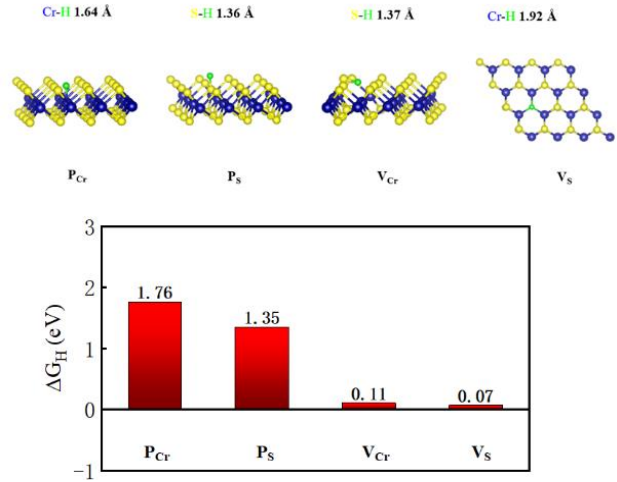

FIG. 3. Upper panel: optimized atomic configurations in turn for $P_{Cr}$, $P_S$, $V_{Cr}$ and $V_S$. Lower panel: their reaction energies for the Volmer reaction.

The formation energy of Cr vacancy is much larger than that for S vacancy, leading to the difficulty in the formation of Cr vacancy in the stoichiometric monolayer. This determines that the $V_{Cr}$ structure cannot be an efficient hydrogen evolution material. Thus, we emphatically discuss the mechanism of HER for the $V_S$ system. $\Delta G_H$ for the Volmer, Heyrovsky, and Tafel reactions are presented in Fig. 4(a). Both first and second Volmer reactions are exothermic due to $\Delta G_H$=0.07 eV and 0.41 eV. $\Delta G_H$ for the Heyrovsky reaction is closer to zero than that for the Tafel reaction. From the perspective of reaction free energy, the Volmer-Heyrovsky reaction seems prone to occur, but such judgment is crude and unreliable, due to the existing of the reaction barrier.

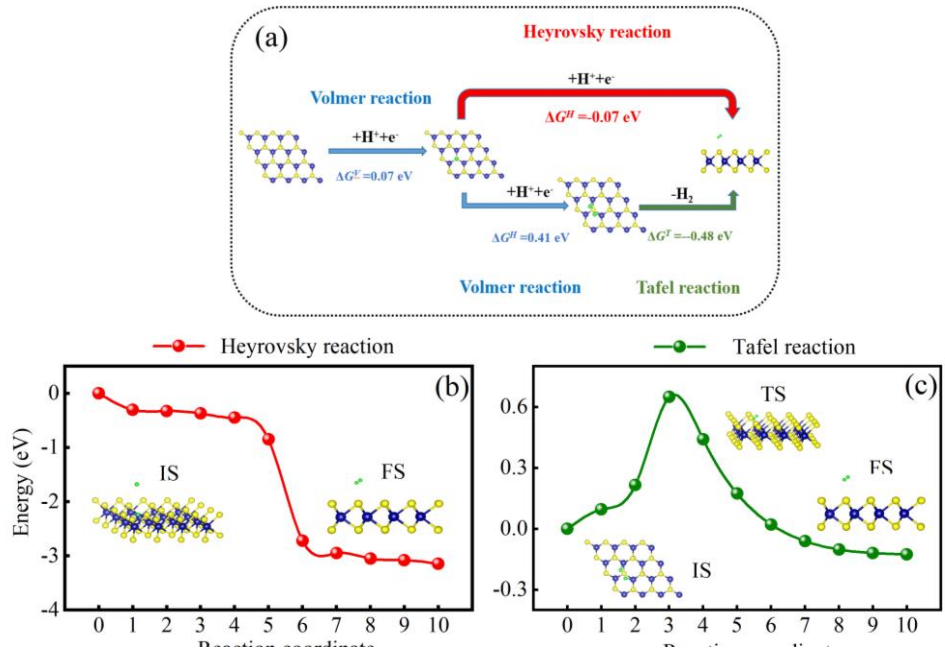

FIG. 4. (a) Sketch of Volmer, Heyrovsky and Talfel reactions on the surface of $V_S$, reaction pathways of (b) the Heyrovsky reaction and (c) the Tafel reaction.

In order to explore the HER mechanism, it is necessary to calculate the reaction pathways. Fig. 4(b) and (c) shows the reaction pathways of the Heyrovsky and Tafel reactions on the surface of the $V_S$ system. For the Tafel reaction, the release of a hydrogen gas molecule through the recombination of $2H^*$ needs to across a transition state with a big energy barrier (0.64 eV), whereas a free proton and an adsorbed H forming a hydrogen gas has nearly no energy barrier for the Heyrovsky reaction. This obviously indicates that the HER on the surface of the $2H-CrS_2$ with S vacancy is prone to the Volmer-Heyrovsky mechanism.

## 4. CONCLUSION

In this work, mechanical, dynamical, and thermal stabilities of $2H-CrS_2$ monolayer are confirmed, and the electronic structures and HER performance and mechanism for the perfect and imperfect $2H-CrS_2$ monolayers are explored. We find the introductions of Cr or S vacancy lead to the reduction of the energy level of d-band center and defect states in the middle of bandgap. The defect states composed of the hybridization of Cr and S can increase electrical conductivity and thus is good for enhanced HER performance. Moreover, our results indicate there are no efficient active sites in the perfect 2H-$CrS_2$ monolayer. However, S vacancies in the imperfect monolayer are excellent active sites. We predicted the 2H-$CrS_2$ monolayer with intrinsic defect S vacancy can be considered as the highly active HER catalyst and its HER is prone to the Volmer−Heyrovsky mechanism.

The data that supports the findings of this study are available within the article [and its supplementary material].

This work is supported by the National Natural Science Foundation of China (Grant No. 11804132).

# SUPPLEMENTARY INFORMATION

# Prediction for structure stability and ultrahigh hydrogen evolution performance of monolayer 2H-CrS$_2$

Feng Sun[1*], Aijun Hong[1], Wenda Zhou[1], Cailei Yuan[1*], Wei Zhang[2]

*Correspondence and requests for materials should be addressed to A.J.H. (email: haj@jxnu.edu.cn) and C.L.Y. (email: clyuan@jxnu.edu.cn).

1 Jiangxi Key Laboratory of Nanomaterials and Sensors, School of Physics, Communication and Electronics, Jiangxi Normal University, Nanchang 330022, China

2 State Key Laboratory of Hydroscience and Engineering, Department of Energy and Power Engineering, Tsinghua University, Beijing 100084, China

TABLE SI. Calculated elastic constants of 2H-CrS$_2$ monolayer (units: GPa)

| $C_{11}$ | $C_{12}$ | $C_{13}$ | $C_{33}$ | $C_{44}$ |
|---|---|---|---|---|
| 64 | 16 | -21 | -6 | 2 |

TABLE SII. Calculated vibrational energy $E_{vib}$ and entropy $S_{vib}$ at 298 K and $E_{ZPE}$ for P$_{Cr}$, P$_S$, V$_{Cr}$ and V$_S$. Experimental vibrational energy $E_{vib}$ and standard entropy $S_{vib}$ for H$_2$ gap molecule from Refs 1 and 2 (units: eV).

|  | P$_{Cr}$ | P$_S$ | V$_{Cr}$ | V$_S$ | H$_2$ |
|---|---|---|---|---|---|
| $E_{vib}$ ($T$=298K) | 0.0990 | 0.2367 | 0.2240 | 0.1587 | 0.28 [1] |
| $TS_{vib}$ ($T$=298K) | 0.0001 | 0.0105 | 0.0117 | 0.0090 | 0.40 [2] |
| $E_{ZEP}$ | 0.0987 | 0.2287 | 0.2153 | 0.1517 | - |